\documentclass[11pt]{article}
\usepackage[utf8]{inputenc}
\pdfoutput=1
\usepackage{amssymb,amsmath,mathrsfs,enumerate}
\usepackage{graphicx,rotate,multicol}
\usepackage{float}
\usepackage{tocloft}
\usepackage{subfig}
\usepackage[margin=10pt,labelfont=bf]{caption}
\usepackage{cite}
\usepackage{soul}
\usepackage{slashed}
\usepackage{multirow}

\usepackage[colorlinks=true,
            linkcolor=blue,
            urlcolor=blue,
            citecolor=teal]{hyperref}

\makeatletter
\let\Hy@linktoc\Hy@linktoc@page
\makeatother

\usepackage{color}
\definecolor{ourcolor}{rgb}{0.7, 0.25, 0.05}
\usepackage{tikz,braket}
\long\def\rpl#1!!#2!!{\textcolor{red}{#1} \textcolor{blue}{#2}}

\def \order(#1){{\mathcal O} \left(#1 \right)}

\evensidemargin=\oddsidemargin
\allowdisplaybreaks

\title{\bf Neutrino Decoherence from Generalised Uncertainty}

\author {\bf Indra Kumar Banerjee,\footnote{indrab@iiserbpr.ac.in} 
\hspace{4pt} Ujjal Kumar Dey\footnote{ujjal@iiserbpr.ac.in} \\[10pt]
\small\em Department of Physical Sciences, Indian Institute of Science Education and Research Berhampur,\\ \small\em Transit Campus, Government ITI, Berhampur 760010, Odisha, India
}

\date{}

\begin{document}
\maketitle

\begin{abstract}
Quantum gravity models predict a minimal measurable length which gives rise to a modification in the uncertainty principle. One of the simplest manifestations of these generalised uncertainty principles is the linear quadratic generalised uncertainty principle which leads to a modified Heisenberg algebra. This can alter the usual von-Neumann evolution of density matrix to a Lindblad-type equation. We show how this can give rise to a decoherence in neutrino propagation in vacuum. The decoherence effects due to the linear quadratic generalised uncertainty principle are extremely minimal and is unlikely to be detectable in the existing or upcoming experimental facilities for any of the natural sources of neutrinos. We also show that, in principle, there can be other variants of generalised uncertainty principle which predicts verifiable decoherence effects for the cosmic neutrino background.
\end{abstract}


\newpage


\section{Introduction}
\label{sec:intro}
The discovery and subsequent verification of the neutrino oscillation phenomenon ushered a new age in the neutrino physics~\cite{Super-Kamiokande:1998kpq, SNO:2001kpb, SNO:2002tuh, T2K:2017hed, NOvA:2017abs}. Current development in the experimental front has already brought neutrino physics into the precision era and thus opened the possibilities to explore physics beyond the standard model (BSM). In fact, the very observation of oscillation phenomenon takes away the masslessness nature of neutrinos, necessitating BSM physics. The mixing between neutrino mass and flavour eigenstates is at the root of the neutrino oscillation. The non-diagonal nature of the charged-current weak interaction, in the mass eigenbasis for both the charged leptons and the neutrinos, is the underlying reason for the aforementioned mixing. More precisely, in the diagonal basis of the charged leptons, the neutrino flavour eigenstates $(\nu_{\alpha},~\alpha = e,\mu,\tau)$ are a linear superposition of the mass eigenstates $(\nu_{i},~i=1,2,3)$, and they are related as $\nu_{\alpha} = U_{\alpha i}\nu_i$, where $U_{\alpha i}$ are the elements of the unitary lepton mixing matrix popularly known as Pontecorvo-Maki-Nakagawa-Sakata (PMNS) matrix. Now, due to the feeble interactions with matter neutrinos can travel enormous distances in isolation from its environment. This makes neutrino oscillation effectively \textit{coherent}, i.e., the wavefunctions of two neutrinos of identical energy travelling along an identical path evolve identically. However, this coherence can be lost over large distances if the neutrinos interact with the environment.  Note that this is different from the wave-packet decoherence that occurs since the packets of the individual propagation eigenstates would get separated in space over long distances~\cite{Chan:2015mca, Akhmedov:2017mcc, Ciuffoli:2022uzf}. Therefore we can dub the former scenario as \textit{environmental decoherence} which is what we are going to focus in this work. The usual matter effects e.g., Mikheev-Smirnov-Wolfenstein (MSW) effect~\cite{Wolfenstein:1977ue, Mikheev:1986wj} and parametric resonance~\cite{Akhmedov:1998ui} etc. are the result from the influence of the multitude of matter particles on propagating neutrinos and therefore these effects consistently work on all traversing neutrinos maintaining the coherence. However, the environmental decoherence can, in principle, alter the known matter effect and can have implications in experimental observations~\cite{Oliveira:2016asf, Coelho:2017byq}.  
Among other things a stochastic environment can be a source of environmental decoherence for neutrinos. In general quantum gravity models predict fluctuating space-time at Planck scales. This space-time ``foam" results in a stochastic medium which can affect the propagation of the particles and effectively result in a decoherence in the neutrinos~\cite{Klapdor-Kleingrothaus:2000kdx, Adler:2000vfa, Barenboim:2006xt, Mavromatos:2006yy, Stuttard:2020qfv,Stuttard:2021uyw}.  To correctly demonstrate the effect of decoherence we need to go beyond the standard quantum mechanical description of neutrino oscillation since due to the system-environment interactions an initial pure quantum state evolves to a mixed state and the ideas from open quantum systems have to be employed.
Quantum decoherence has been a subject of interest since the 1970s~\cite{Zeh:1970zz}. It was proposed as an explanation of the transition from quantum to classical in case of macroscopic behaviour of a system. It explains how a purely coherent quantum system interacts with the environment and loses its characteristic quantum feature, i.e., the wavefuncions' ability to interfere with themselves. 
In this process the population of a quantum system changes due to the entanglement with the environment which can be a thermal bath or due to some other phenomenon~\cite{Zanardi:1997vn}. The evolution of the density matrix of a state due to decoherence is governed by  Gorini-Kossakowski-Sudarshan-Lindblad master equation instead of the von-Neumann equation~\cite{Gorini:1975nb, Lindblad:1975ef}. Using the framework of open quantum systems neutrino decoherence has been extensively studied from phenomenological and experimental point of view~\cite{Benatti:2000ph, Ohlsson:2000mj,  Lisi:2000zt,Klapdor-Kleingrothaus:2000kdx, Adler:2000vfa, Mavromatos:2007zz, Mavromatos:2007hv, Fogli:2007tx, Guzzo:2014jbp,  BalieiroGomes:2018gtd,Capolupo:2018hrp, Stuttard:2020qfv,  Gomes:2020muc, deGouvea:2020hfl, Cheng:2022lys, Stuttard:2021uyw}.
In this paper we explore the possibility neutrino decoherence that can arise from a different manifestation of quantum gravity ideas. The most promising candidates of quantum gravity, namely, string theory, loop quantum gravity, modified dispersion relations and doubly special relativity etc. all predicts a minimum measurable length approximately of the order of the Planck length $l_p \sim 10^{-35}\text{ m}$. 
A constraint of this kind modifies the Heisenberg uncertainty principle to \textit{generalised uncertainty principle} (GUP)~\cite{Kempf:1994su, Adler:2001vs, Ali:2009zq, Das:2010zf, Chen:2014jwq, Chen:2014bva, Scardigli:2016pjs, Kanazawa:2019llj, Luciano:2019mrz}. 
Some of the interesting phenomenological and theoretical implications of GUP can be found in~\cite{KalyanaRama:2001xd, Hossenfelder:2003jz, Sprenger:2012uc, Pramanik:2014mma, Faizal:2014mfa, Ali:2015ola, Faizal:2016zlo, Masood:2016wma, Lake:2017uzd, Abhishek:2022ycj}.
The manifestation of the generalisation is contained in the deformed canonical commutation relations and they in turn gives rise to modified equations for the evolution of a quantum state which are Lindblad-type master equations~\cite{Petruzziello:2020wkd, Al-Nasrallah:2021zie} which will determine the time evolution of the neutrino state and ultimately give rise to the decoherence. The effect of decoherence is in general parametrised by the energy-dependent decoherence parameter $\Gamma$. Due to the ambiguity in the theory of quantum gravity sometimes an empirical approach is taken and a power law energy dependence of $\Gamma$ is assumed~\cite{Lisi:2000zt, Fogli:2007tx, Guzzo:2014jbp},
\begin{align}
\Gamma(E)=\Gamma(E_0)\left(\frac{E}{E_0}\right)^n. 
\end{align}
In some studies, experimental observations have been used to determine the validity of different values of $n$ in a model independent way~\cite{Klapdor-Kleingrothaus:2000kdx, Adler:2000vfa, Stuttard:2020qfv,Stuttard:2021uyw}. In this study, we take a model-dependent approach to use the Lindblad-type master equation arising from a GUP to investigate the decoherence effect on the neutrino sector.
The paper is organised as follows: in section~\ref{sec:nuOsc} we take a short recap on the conventional treatment of the neutrino oscillation, in section~\ref{sec:gupLindblad} we discuss the generalised uncertainty principle and the corresponding deformed canonical commutation relation used in this study. In section~\ref{sec:nuDecoh} decoherence effects on neutrino due to this model has been explained. Possible extensions of the current scenario are discussed in section~\ref{sec:extn}. We explain the results in section~\ref{sec:results} and finally we summarize and conclude.

\section{Conventional Treatment of Neutrino Oscillation}
\label{sec:nuOsc}
In this section we briefly describe the conventional treatment of neutrino oscillation to set our notations and conventions. 

In general, neutrino flavour eigenstates can be expressed in terms of the neutrino mass eigenstates with the PMNS mixing matrix as,
\begin{equation}
\label{eq:nuFlav}
\ket{\nu_{\alpha}}=\sum_{a=1}^{n}U_{\alpha a}^{*}\ket{\nu_a},
\end{equation} 
where $n$ is the number of neutrino mass eigenstates and $U$ is the PMNS matrix. Clearly, for the two neutrino case, $n=2$ and $\alpha = e, \mu$, and for the three neutrino case $n=3$ and $\alpha = e,\mu,\tau$. In general the neutrino oscillation probabilities can be given by the  well-known expression~\cite{Giunti:2007ry}, 
\begin{align}
P_{\alpha\beta}(L,E) = \delta_{\alpha\beta} 
    & -4\sum_{a > b}
        \text{Re}(U^{*}_{\alpha a}U_{\beta a}U_{\alpha b}
        U^{*}_{\beta b})\sin^{2}
        \frac{\Delta m^{2}_{ab}L}{4E} \nonumber \\
    & -2\sum_{a > b}\text{Im}
       (U^{*}_{\alpha a}U_{\beta a}U_{\alpha b}
       U^{*}_{\beta b})\sin \frac{\Delta m^{2}_{ab}L}{2E},
\end{align} 
where $\alpha, \beta = e, \mu, \tau, \ldots$, $L$ is the neutrino path length, $E$ is the neutrino energy, $\Delta m^{2}_{ab}=m_a^2-m_b^2$, $a,b = 1,2,3,\ldots$ is the mass squared difference between the neutrino mass eigenstates $|\nu_a \rangle$ and $|\nu_b \rangle$ with masses $m_a$ and $m_b$ respectively. The oscillation probability for $\alpha \neq \beta$ is termed as \textit{transition probability} and the $\alpha = \beta$ case is known as \textit{survival probability}. 
In the two-flavour case, the mixing matrix $U$ takes the form 
\begin{equation}
\label{eq:2dPMNS}
U=\begin{pmatrix}
\cos\theta & \sin\theta\\
-\sin\theta & \cos\theta
\end{pmatrix}.
\end{equation} 
%
However, in the realistic three-flavour case the mixing matrix $U$ is the famous PMNS matrix which is given by~\cite{Workman:2022ynf}, 

\begin{equation}
U = \begin{pmatrix}
c_{12}c_{13} & s_{12}c_{13} & s_{13}e^{-i\delta} \\ 
-s_{12}c_{23}-c_{12}s_{23}s_{13}e^{i\delta} 
& c_{12}c_{23}-s_{12}s_{23}s_{13}e^{i\delta} 
& s_{23}c_{13} \\ 
s_{12}s_{23}-c_{12}c_{23}s_{13}e^{i\delta} 
& -c_{12}s_{23}-s_{12}c_{23}s_{13}e^{i\delta} 
& c_{23}c_{13}
\end{pmatrix},
\end{equation}
where the $c_{ij} = \cos\theta_{ij}$ and the $s_{ij} = \sin\theta_{ij}$ are used to denote the cosine and sine of the mixing angles $\theta_{ij}$ respectively and $\delta$ is the CP-violating phase. Conventionally, they are taken to be in the ranges $\theta_{ij}\in [0,\pi/2]$ and $\delta \in [0,2\pi]$. Here we do not mention the possibility of the presence of Majorana phases in this $3\times 3$ mixing matrix, since in the oscillation phenomena they play no roles. For our numerical analysis we will use the experimentally obtained best-fit values from the NuFIT collaboration~\cite{Esteban:2020cvm}, for the normal ordering
\begin{align}
\Delta m_{21}^2 &=7.42^{+0.21}_{-0.20}\times 10^{-5}\text{ eV}^2,\\
\Delta m_{31}^2 &=2.514^{+0.028}_{-0.027}\times 10^{-3}\text{ eV}^2,\\ 
\delta &=195^{\circ}~^{+51^{\circ}}_{-25^{\circ}},
\label{NOParam}
\end{align}
and for the inverted ordering.
\begin{align}
\Delta m_{21}^2 &=7.42^{+0.21}_{-0.20}\times 10^{-5}\text{ eV}^2,\\
\Delta m_{23}^2 &=2.517^{+0.026}_{-0.028}\times 10^{-3}\text{ eV}^2,\\ 
\delta &=197^{\circ}~^{+27^{\circ}}_{-24^{\circ}}.
\label{IOParam}
\end{align}
Moreover, for our purposes we will consider just the central values of these parameter.
In passing we note that the transition probability is sometimes taken as Gaussian average. Actually, due to the lack of sharp energy or well-defined propagation length, a Gaussian average over the $L/E$ dependence and other uncertainties are taken. The averaged transition probability takes the form,
\begin{align}
\label{eq:oscProbGauss}
P_{\alpha\beta}(L,E) = \delta_{\alpha\beta}
    &-2\sum_{a=1}^{n}\sum_{b=1}^{n}\text{Re}(U^{*}_{\alpha a}
      U_{\beta a}U_{\alpha b}U^{*}_{\beta b})
      \left(1-\cos \frac{\Delta m^{2}_{ab}L}{2E}
      e^{-2\sigma^2(\Delta m^2_{ab})^2}\right)\nonumber \\
    &-2\sum_{a=1}^{n}\sum_{b=1}^{n}\text{Im}(U^{*}_{\alpha a}
      U_{\beta a}U_{\alpha b}U^{*}_{\beta b})
      \sin \frac{\Delta m^{2}_{ab}L}{2E}
      e^{-2\sigma^2(\Delta m^2_{ab})^2},
\end{align}
where $\sigma$ is the damping parameter which is responsible for the damping of the neutrino oscillation probabilities. It is to be noted that all the variables in the above equation e.g., $P_{\alpha \beta}$, $L$, $E$ are now averages of the transitions probability, neutrino propagation length and the neutrino energy respectively. The equivalence between this kind of averaging and the neutrino decoherence was shown for the first time in~\cite{Ohlsson:2000mj}. In the next sections we will show this in the present context too.


\section{Generalised Uncertainty Principle and Lindblad Master Equation}
\label{sec:gupLindblad}
In this section we discuss the generalised uncertainty principle (GUP) and set-up the Lindblad equation relevant for our studies.  
\subsection{GUP and DCCR}
\label{sbsec:gupdccr}
The Heisenberg's Uncertainty principle, $\Delta x\Delta p \ge \hbar/2$ is a direct consequence of the non-commutativity of the position and the momentum operators, i.e. $[x_i,p_j]=i\hbar\delta_{ij}$. As the various theories of quantum gravity suggest, this commutation relation is modified, i.e. we get a deformed canonical commutation relation (DCCR) in order to accommodate the prediction of a minimum measurable length in such theories. The modification presents itself as higher order terms of the momentum observable in the right-hand-side of the commutator relation. An example of such a DCCR, which is motivated by string theory, can be~\cite{Petruzziello:2020wkd},
\begin{equation}
[x_i,p_j] = i\hbar[\delta_{ij}+\beta\delta_{ij}p^2+2\beta p_ip_j],
\end{equation} 
where $\beta = \beta_{0} l_{P}^{2}/\hbar^2$ with $l_P$ being the Planck length and $\beta_0$ is some constant. DCCRs of this form can give rise to the GUP, 
\begin{equation}
\Delta x\Delta p \ge \dfrac{\hbar}{2}(1+\beta\Delta p^2) .
\end{equation} 
A more general approach, supported by doubly special relativity, has been investigated to construct a GUP which contains both linear and quadratic terms of momentum observable in the right hand side (LQGUP). This can be of the form~\cite{Al-Nasrallah:2021zie},
\begin{equation}
\label{eq:lqgup}
\Delta x\Delta p \ge \dfrac{\hbar}{2}(1-2\alpha \langle p \rangle + \alpha^2 \langle p \rangle^2),
\end{equation} 
which is resulted in from the following DCCR, 
\begin{equation}
\label{eq:lqcommut}
[x_i,p_j]=i\hbar\left[\delta_{ij}-\alpha \left(p\delta_{ij}+\dfrac{p_ip_j}{p}\right)+\alpha^2(p^2\delta_{ij}+3p_ip_j)\right].
\end{equation} 
Here, the parameter $\alpha$ is defined as $\alpha = \alpha_0 l_P/\hbar$. Here $\alpha$ is the GUP parameter and it is often treated as a Gaussian white noise for simplicity to comply with the existence of a fluctuating minimum length scale as predicted by the theories as mentioned before. The quantity $\alpha_0$ is often taken to be the order of unity.

\subsection{Generalisation to Lindblad equation}
\label{sbsec:lindblad}
Using the LQGUP mentioned in the previous section, Eq.~\eqref{eq:lqgup}, a Lindbland-type master equation can be constructed which determines the evolution of the quantum system corresponding to the decoherence created by the GUP which in turn arises from the quantum gravity aspect of the theory. The DCCR gives rise to modified momentum which satisfies the Heisenberg's Uncertainty Principle~\cite{Al-Nasrallah:2021zie}. This means that taking cue from Eq.~\eqref{eq:lqcommut} we can write the operators $x_i$ and $p_j$ in the following form,   
\begin{equation}
\label{eq:modmomrel}
x_i=x_{0i},~~p_i=p_{0i}(1-\alpha p_0+2\alpha^2p_{0}^{2}),
\end{equation}
where $x_{0i}$ and $p_{0i}$ satisfy the conventional relation $[x_{0i}, p_{0j}] = i\delta_{ij}$. In addition the other commutation relations remain intact, i.e., $[x_i, x_j] = 0 = [p_i, p_j]$. Note that from here onward we will use the natural unit $\hbar = 1$.
The Lindbland-type master equation from the LQGUP depends on the Hamiltonian of the system. For example, in~\cite{Al-Nasrallah:2021zie}, the Hamiltonian was that of a non-relativistic particle. Here, since we will be considering neutrinos, we take the Hamiltonian of an ultra-relativistic free particle. The Schrödinger's equation in this case becomes,
\begin{equation}
\label{eq:schrodultra}
i\dfrac{d\psi}{dt}=\left(\left|\vec{p}\right|+\dfrac{m^2}{2\left|\vec{p}\right|}\right)\psi
\end{equation}
Using the modified relations in Eq.~\eqref{eq:modmomrel}, we get
\begin{equation}
\label{eq:modschrod}
i\dfrac{d\psi}{dt}=\left(p_{0}(1-\alpha p_0+2\alpha^2p_{0}^{2})+\dfrac{m^2}{2p_{0}(1-\alpha p_0+2\alpha^2p_{0}^{2})}\right)\psi
\end{equation}
Now we expand this and we take into account terms upto first order in $\alpha$ to get,
\begin{equation}
\label{eq:modschrod2}
i\dfrac{d\psi}{dt}=\left(H_0 + H_I+\mathcal{O}(\alpha^2)\right)\psi
\end{equation}
where $H_0=p_0+\dfrac{m^2}{2p_0}$ and $H_I=-\alpha\left(p_0^2-\dfrac{m^2}{2}\right)$.
In this case, the Lindblad-type equation can be derived to be,
\begin{equation}
\label{eq:modlindblad}
\dot{\rho}=-i[H,\rho(t)] - 
\gamma \left[\left(H^{2}-\dfrac{3}{2}m^2\right),\left[\left(H^{2}-\dfrac{3}{2}m^2\right),\rho(t)\right]\right],
\end{equation} 
where $\gamma = t_{P}l_{P}^{2}$. Here the second term in the right hand side (RHS) of the equation is the dissipator or the Lindblad-type operator. We can express the above equation in a simpler way using the Lindblad type operators, $D=\sqrt{\gamma}\left(H^2-\dfrac{3}{2}m^2\right)$.
\begin{equation}
\label{eq:modlindbladgen}
\dot{\rho}=-i[H,\rho(t)]-[D,[D,\rho(t)]].
\end{equation} 
In the next sections, we use this Lindblad-type master equation to investigate the decoherence in the neutrino flavour oscillation in vacuum.

\section{Neutrino Decoherence}
\label{sec:nuDecoh}

\subsection{Two flavour case}
\label{sbsec:2flavDecoh}
Neutrino flavour transitions through the decoherence mechanisms have been studied for Lindblad operators that are linearly dependent of the hamiltonian and they produce transition probabilities similar to that of Eq.~\eqref{eq:oscProbGauss}~\cite{Ohlsson:2000mj}. In this case we use the Lindblads which are not linearly dependent on the Hamiltonian as shown in the Eqs.~\eqref{eq:modlindblad} and~\eqref{eq:modlindbladgen}. For the two flavour case, the effective Hamiltonian in the mass eigenstate basis can be written as,
\begin{equation}
H = \frac{1}{4E}
 \begin{pmatrix}
 -\Delta m^2 & 0 \\
   0 & \Delta m^2
 \end{pmatrix}
 = \frac{\Delta m^2}{4E}
 \begin{pmatrix}
 -1 & 0 \\
  0 & 1
 \end{pmatrix}
  = \omega_2
 \begin{pmatrix}
 -1 & 0 \\
  0 & 1
 \end{pmatrix},
\end{equation}
where $\Delta m^2 \equiv m_2^2 - m_1^2$, $E$ is the energy of the neutrinos and $\omega_2=\dfrac{\Delta m^2}{4E}$.
In this case the Lindblad operators takes the form,
\begin{equation}
\label{eq:LindOp}
D_a = \sqrt{\gamma}(\omega_2^2-\frac{3}{2}m_a^2)
     \begin{pmatrix}
     1 & 0 \\0 & 1
     \end{pmatrix}
    = \lambda_a 
      \begin{pmatrix}
      1 & 0 \\
      0 & 1
      \end{pmatrix},
\end{equation} 
where $\lambda_a=\sqrt{\gamma}\left(\omega_2^2-\frac{3}{2}m_a^2\right)$. Here the index $a$ represents different mass eigenstates. It is evident that since the $D_a$ terms are proportional to the identity matrix, there will be no decoherence effect since the second term in the RHS of Eq.~\eqref{eq:modlindbladgen} is vanishing. Nevertheless, below we show this explicitly and in the process lay down the steps of the oscillation probability calculation that will be used in the three flavour case as well. 
The relevant density matrix $\rho$ whose elements are functions of time and which follows the condition $\text{Tr}(\rho)=1$ can be written in the following form,
\begin{equation}
\label{eq:2drho}
\rho = \frac12 
       \begin{pmatrix}
       1+a & p+iq \\
       p-iq & 1-a
       \end{pmatrix},
\end{equation} 
where $a,p,q$ are real valued functions of time, i.e., $a=a(t)$, $p=p(t)$ and $q=q(t)$. Inserting Eqs.~\eqref{eq:LindOp} and~\eqref{eq:2drho} into Eq.~\eqref{eq:modlindbladgen} we get the Lindblad-type master equation for the evolution of the density matrix in the two-flavour case as,
\begin{align}
\begin{pmatrix}
\dot{a} & \dot{p}+i\dot{q} \\
\dot{p}-i\dot{q} & -\dot{a}
\end{pmatrix} 
= \omega_2
 \begin{pmatrix}
 0 & 2ip-2q \\
 -2ip-2q & 0
 \end{pmatrix} 
 - \lambda^2
 \begin{pmatrix}
 0 & 0 \\
 0 & 0
 \end{pmatrix},
\end{align} 
where $\lambda^2=\sum_{a=1}^{2}\lambda^2_a$.
From the above equation, the solutions for the parameters can be obtained as,
\begin{subequations}
\label{eq:apq}
\begin{align}
a(t)&=a(0),
\\p(t)&=\left[p(0)\text{cos}(2\omega_2t) + q(0)\text{sin}(2\omega_2t)\right],
\\q(t)&=\left[-p(0)\text{sin}(2\omega_2t) + q(0)\text{cos}(2\omega_2t)\right].
\end{align} 
\end{subequations}
We can now take mass eigenstates to be,
\begin{align}
\ket{\nu_1} = 
\begin{pmatrix}1 \\ 0 \end{pmatrix}, 
~~\text{and}~~
\ket{\nu_2} = 
\begin{pmatrix}0 \\ 1 \end{pmatrix}, 
\end{align}
and following Eqs.~\eqref{eq:nuFlav} and~\eqref{eq:2dPMNS}, the flavour eigenstates can be written as,
\begin{align}
\ket{\nu_e} = 
\begin{pmatrix}
\cos\theta \\ 
\sin\theta
\end{pmatrix},~~ \text{and}~~
\ket{\nu_\mu} = 
\begin{pmatrix}
-\sin\theta \\ 
\cos\theta
\end{pmatrix},
\end{align}
for electron neutrino and muon neutrino, respectively. Now, if the system is initially in the electron neutrino state, then the initial density matrix is, 
\begin{equation}
\rho_{e}(0) = \ket{\nu_e}\bra{\nu_e}
            = \begin{pmatrix}
              \cos^2\theta & \sin\theta \cos\theta \\
              \sin\theta \cos\theta & \cos^2\theta
              \end{pmatrix}.
\end{equation}
From Eq.~\eqref{eq:2drho} the initial conditions for the parameters can be determined as, $a(0)=\text{cos}2\theta$, $p(0)=\text{sin}2\theta$ and $q(0)=0$.
Therefore, the neutrino transition probabilities with the decoherence effects are,
\begin{subequations}
\begin{align}
\label{eq:PeeDecoh}
P_{ee}(t) &= \text{Tr}(\ket{\nu_e}\bra{\nu_e}\rho(t))
          = \frac{1}{2}\left[1+\sin2\theta p(t)
            +\cos2\theta a(t)\right], \\
\label{eq:PemuDecoh}            
P_{e\mu}(t) &= \text{Tr}(\ket{\nu_\mu}\bra{\nu_\mu}\rho(t))
 = \frac{1}{2}\left[1-\sin2\theta p(t)-\cos2\theta a(t)\right].          
\end{align}
\end{subequations}
Using Eqs.~\eqref{eq:apq} in Eq.~\eqref{eq:PemuDecoh}, we get transition probability as,
\begin{equation}
P_{e\mu}(t) = \frac{1}{2}\sin^2 2\theta
  \left[1-\cos \left(
  \frac{\Delta m^2t}{2E}\right)\right].
\end{equation} 
Assuming $t\sim L$ in the above equation we get,
\begin{equation}
P_{e\mu}(L)=\dfrac{1}{2}\sin^2 2\theta\left[
1-\cos \left(
\frac{\Delta m^2L}{2E}\right)\right].
\end{equation}
As we can see from the above expression that for the two flavour case, the Lindblad-type equation does not give rise to any decoherence upto the first order of $\alpha$ in the Hamiltonian. This, as expected, is due to the fact that the Lindblad operators are proportional to the identity matrix. Hence, the dissipation term gets cancelled.

\subsection{Three flavour case}
\label{sbsec:3flav}

In the three flavour case we can have two scenarios, namely the normal and inverted hierarchy between the eigenstates in the mass basis. In the normal mass hierarchy we assume, 
\begin{equation}
m_3^2 \gg m_2^2 > m_1^2.
\end{equation}
The Hamiltonian in the mass eigenbasis can be given by,
\begin{equation}
\label{eq:Ham3by3}
H = \frac{1}{2E}
\begin{pmatrix} 
0 & 0 & 0 \\ 
0 & \Delta m_{21}^2 & 0 \\ 
0 & 0 & \Delta m_{31}^2
\end{pmatrix}.
\end{equation}
In normal mass hierarchy, $\Delta m_{21}^2,\Delta m_{31}^2>0$, Hence, the Lindblad operators take the form 
\begin{equation}
\label{eq:Lind3by3}
D_a = \sqrt{\gamma}
\begin{pmatrix} 
-\frac{3}{2}m_a^2 & 0 & 0 \\ 
0 & \omega_3^2(\Delta m_{21}^2)^{2}-\frac{3}{2}m_a^2 & 0 \\ 
0 & 0 & \omega_3^2(\Delta m_{31}^2)^{2}-\frac{3}{2}m_a^2
\end{pmatrix} 
\end{equation} 
where $\omega_3 = 1/2E$.
Again, similar to Eq.~\eqref{eq:2drho} we can define the most general density matrix for the three flavour case as,
\begin{equation}
\label{eq:3drho}
\rho = \begin{pmatrix}
a & p+iq & f+ig \\ 
p-iq & b & x+iy \\ 
f-ig & x-iy & c
\end{pmatrix}.
\end{equation}
Here, $a,b,c,p,q,f,g,x,y$ are all real-valued functions of time. In order to maintain the condition $\text{Tr}(\rho)=1$, the constraint on these variables is ,
\begin{equation}
a+b+c = 1.
\end{equation}
Now, using Eqs.~\eqref{eq:Ham3by3}-\eqref{eq:3drho} in Eq.~\eqref{eq:modlindbladgen}, we get,
\begin{equation}
\label{eq:rhodot3by3}
\begin{pmatrix}
\dot{a} & \dot{p}+i\dot{q} & \dot{f}+i\dot{g} \\ 
\dot{p}-i\dot{q} & \dot{b} & \dot{x}+i\dot{y} \\ 
\dot{f}-i\dot{g} & \dot{x}-i\dot{y} & \dot{c}
\end{pmatrix} 
= \begin{pmatrix}
A_{11} & A_{12} & A_{13} \\ 
A_{12}^* & A_{22} & A_{23} \\ 
A_{13}^* & A_{23}^* & A_{33}
\end{pmatrix},
\end{equation}
where 
\begin{subequations}
\begin{align}
A_{11} &= A_{22} = A_{33} = 0,
\label{eq:AMatDiagElem}
\\
A_{12} &= \left[-\omega_3\Delta m_{21}^2q-3\gamma\omega_3^4(\Delta m_{21}^2)^4p\right]
+ i\left[\omega_3\Delta m_{21}^2p-3\gamma\omega_3^4(\Delta m_{21}^2)^4q\right],\\
A_{13} &= \left[-\omega_3\Delta m_{31}^2g-3\gamma\omega_3^4(\Delta m_{31}^2)^4f\right]
+ i\left[\omega_3\Delta m_{31}^2f-3\gamma\omega_3^4(\Delta m_{31}^2)^4g\right],\\
A_{23} &= \left[-\omega_3\Delta m_{32}^2y-3\gamma\omega_3^4(\Delta m_{32}^2)^2(\Delta m_{31}^2+\Delta m_{21}^2)^2x\right]\\ \nonumber
&~~~~~~+ i\left[\omega_3\Delta m_{32}^2x-3\gamma\omega_3^4(\Delta m_{32}^2)^2(\Delta m_{31}^2+\Delta m_{21}^2)^2y\right].
\end{align}
\end{subequations}
Note, from Eqs.~\eqref{eq:rhodot3by3} and~\eqref{eq:AMatDiagElem} we that the diagonal elements of the density matrix $\rho$ are time-independent, i.e., 
\begin{equation}
\label{eq:diagabc}
a(t)=a(0),~b(t)=b(0),~c(t)=c(0).
\end{equation}
The other elements are calculated to be,
\begin{align}
p(t) &= e^{-3\gamma\omega_3^4(\Delta m_{21}^2)^4t}
       \left[p(0)\cos(\omega_3\Delta m_{21}^2t) 
        -q(0)\sin(\omega_3\Delta m_{21}^2t)\right],\\
q(t) &= e^{-3\gamma\omega_3^4(\Delta m_{21}^2)^4t}
       \left[q(0)\cos(\omega_3\Delta m_{21}^2t)
       +p(0)\sin(\omega_3\Delta m_{21}^2t)\right],\\
f(t) &= e^{-3\gamma\omega_3^4(\Delta m_{31}^2)^4t}
       \left[f(0)\cos(\omega_3\Delta m_{31}^2t)
       -g(0)\sin(\omega_3\Delta m_{31}^2t)\right],\\
g(t) &= e^{-3\gamma\omega_3^4(\Delta m_{31}^2)^4t}
       \left[g(0)\cos(\omega_3\Delta m_{31}^2t)
       +f(0)\sin(\omega_3\Delta m_{31}^2t)\right],\\
x(t) &= \exp[-3\gamma\omega_3^4(\Delta m_{32}^2)^2(\Delta m_{31}^2+\Delta m_{21}^2)^2t]
       \left[x(0)\cos(\omega_3\Delta m_{32}^2t)
       -y(0)\sin(\omega_3\Delta m_{32}^2t)\right],\\
\label{eq:yt}       
y(t) &= \exp[-3\gamma\omega_3^4(\Delta m_{32}^2)^2(\Delta m_{31}^2+\Delta m_{21}^2)^2t]
       \left[y(0)\cos(\omega_3\Delta m_{32}^2t)
       +x(0)\sin(\omega_3\Delta m_{32}^2t)\right],
\end{align}
%
Using a method similar to that of in subsection~\ref{sbsec:2flavDecoh}, the transition probabilities of the neutrino flavours can be calculated from Eqs. \eqref{eq:3drho}, \eqref{eq:diagabc}-\eqref{eq:yt}. For example, the survival and transition probabilities for the electron and muon flavours, with the decoherence effects can now be written as,
\begin{align}
\label{eq:Pee3by3}
P_{ee} = 1&-2c_{12}^2s_{12}^2c_{13}^4-2c_{12}^2c_{13}^2s_{13}^2-2s_{12}^2c_{13}^2s_{13}^2 + 2\cos\left(\frac{\Delta m^2_{21}t}{2E}\right)e^{-3\gamma\omega_3^4(\Delta m_{21}^2)^4t}c_{12}^2s_{12}^2c_{13}^4 \nonumber \\
&+2\cos\left(\frac{\Delta m^2_{31}t}{2E}\right)e^{-3\gamma\omega_3^4(\Delta m_{31}^2)^4t}c_{12}^2c_{13}^2s_{13}^2\nonumber
\\&+2\cos\left(\frac{\Delta m^2_{32}t}{2E}\right)e^{-3\gamma\omega_3^4(\Delta m_{32}^2)^2(\Delta m_{31}^2+\Delta m_{21}^2)^2t}s_{12}^2c_{13}^2s_{13}^2 ,
\end{align}
\begin{align}
\label{eq:Pemu3by3} 
P_{e\mu} &=2c_{12}^2 s_{12}^2 c_{13}^2(c_{23}^2 -s_{13}^2 s_{23}^2) - 2s_{23}^2c_{13}^2s_{13}^2 - 2\cos\delta\left(c_{12}s_{12}^3c_{23}s_{23}c_{13}^2s_{13} - c_{12}^3s_{12}c_{23}c_{13}s_{13}\right) \nonumber  \\ 
&-4\left\lbrace c_{12}^2s_{12}^2c_{13}^2(c_{23}^2 - s_{23}^2 s_{13}^2) - s_{13}c_{23}c_{13}^2\cos\delta (c_{12}s_{12}^3s_{23}s_{23} - c_{12}^3s_{12}) \right\rbrace \nonumber \\
&\times \cos\left(\frac{\Delta m^2_{21}t}{2E}\right)e^{-3\gamma\omega_3^4(\Delta m_{21}^2)^4t}\nonumber \\
&+ 2\left(c_{12}s_{12}c_{23}s_{23}c_{13}^2s_{13}\cos\delta + c_{12}^2s_{23}^2c_{13}^2s_{13}^2\right)\cos\left(\frac{\Delta m^2_{31}t}{2E}\right)e^{-3\gamma\omega_3^4(\Delta m_{31}^2)^4t} \nonumber \\
&+ 4c_{13}^2\left(s_{12}^2s_{23}^2s_{13}^2 -c_{12}s_{12}c_{23}s_{23}s_{13}\cos \delta\right)\cos\left(\frac{\Delta m^2_{32}t}{2E}\right)e^{-3\gamma\omega_3^4(\Delta m_{32}^2)^2(\Delta m_{31}^2+\Delta m_{21}^2)^2t} \nonumber \\
&+2c_{13}^2\left(c_{12}s_{12}^3c_{23}s_{23}s_{13}\sin\delta + c_{12}^3s_{12}c_{23}s_{23}s_{13}\sin \delta\right)\sin\left(\frac{\Delta m^2_{21}t}{2E}\right)e^{-3\gamma\omega_3^4(\Delta m_{21}^2)^4t} \nonumber \\
&+ 2c_{12}s_{12}c_{23}s_{23}c_{13}^2s_{13}\sin\delta~\sin\left(\frac{\Delta m^2_{31}t}{2E}\right)e^{-3\gamma\omega_3^4(\Delta m_{31}^2)^4t} \nonumber \\
&+ 2c_{12}s_{12}c_{23}s_{23}c_{13}^2s_{13}\sin\delta~\sin\left(\dfrac{\Delta m^2_{32}t}{2E}\right)e^{-3\gamma\omega_3^4(\Delta m_{32}^2)^2(\Delta m_{31}^2+\Delta m_{21}^2)^2t}.
\end{align}

Here, we can use $t\sim L$ in Eqs.~\eqref{eq:Pee3by3} and~\eqref{eq:Pemu3by3} and can very easily identify that they follow the general form mentioned in Eq.~\eqref{eq:oscProbGauss}. The decoherence effects manifest themselves in the exponentially decaying terms. The survival and transition probabilities for the other generations can also be obtained following the similar prescription.

For the inverted hierarchy case, the mass ordering is different than that of the normal mass hierarchy, 
i.e., $m_2^2 > m_1^2 \gg m_3^2$. In this case, the Hamiltonian takes the form
\begin{equation}
H = \frac{1}{2E}
\begin{pmatrix} 
0 & 0 & 0 \\ 
0 & \Delta m_{21}^2 & 0 \\ 
0 & 0 & -\Delta m_{13}^2
\end{pmatrix}.
\end{equation}
Therefore, the Lindblad operators take the form, 
\begin{equation}
\label{eq:DaInv}
D_a = \sqrt{\gamma}
\begin{pmatrix} 
-\frac{3}{2}m_a^2 & 0 & 0 \\ 
0 & \omega_3^2(\Delta m_{21}^2)^{2}-\frac{3}{2}m_a^2 & 0 \\ 
0 & 0 & \omega_3^2(\Delta m_{13}^2)^{2}-\frac{3}{2}m_a^2
\end{pmatrix}
\end{equation} 
From Eqs.~\eqref{eq:Lind3by3} and \eqref{eq:DaInv} it is evident that the Lindblad operators are the same for both normal and inverted mass hierarchy. Therefore, the decoherence effect in the survival and transition probabilities will be similar in normal and inverted hierarchies.

\section{Possible extension of the model}
\label{sec:extn}
Normally in the study of neutrino decoherence, the decoherence parameters, i.e. the argument of the exponentially decaying terms in the transition and survival probabilities are represented as power law of the energy, i.e., $\Gamma \propto E^{n}$~\cite{Klapdor-Kleingrothaus:2000kdx, Stuttard:2020qfv,Stuttard:2021uyw}. In case of the LQGUP the decoherence parameters, $\Gamma \propto E^{-4}$ as can be seen from Eqs.~\eqref{eq:Pee3by3} and \eqref{eq:Pemu3by3}. Motivated by different quantum gravity scenarios there can be other forms of GUPs as well. In principle, they can give rise to different power law relations for the decoherence parameters. In this section, we show how this can be achieved and what modification will be there in the Heisenberg algebra for such a scenario.
For this section, we try to construct a Lindblad-type equation which generates a neutrino decoherence with decoherence parameters which are $\propto E^{-2}$ and then we try to generate the modified momentum relationship corresponding to the Heisenberg algebra. 
In Eq.~\eqref{eq:Ham3by3}, the Hamiltonian for the neutrino is given in the mass eigenstate. Therefore, in order to generate a $n=-2$ model for the deocherence, we need the Lindblad operators $D_a \propto H$. Hence, the form of the modified Lindblad-type equation in this case is,
\begin{align}
\label{eq:modifLind}
\dot{\rho} =-i[H,\rho(t)]-A[H,[H,\rho(t)]],
\end{align}
where $A$ depends on the GUP parameter $\alpha$ as given in subsection~\ref{sbsec:gupdccr}. Now, in order to achieve such an equation using the treatments given in~\cite{Petruzziello:2020wkd, Al-Nasrallah:2021zie}, the perturbation Hamiltonian should be of the form,
\begin{align}
    H_I &= -\alpha^{1/6}H_0,
\end{align}
where $p_0^2 = \sum_{i=1}^3 p_{0i}p_{0i}$ and $p_i$ and $p_{0i}$ are related as, $p_i = p_{0i}(1+\alpha^{1/6}+\alpha^{1/12}+B(\alpha, p_0))$. Here, the function $B(\alpha,p_0)$ contains all other terms in $\alpha$ and $p_0$. The particular form of $B(\alpha,p_0)$ depends on the GUP that generates this modified Heisenberg algebra. This can lead to the Lindblad-type Eq.~\eqref{eq:modifLind}, in which case,
\begin{align}
    A=(t_pl_p^2)^{1/3}.
\end{align}
Similarly, a dependence of the form $n=-3$ can arise from a relation $p_i = p_{0i}(1+\alpha^{1/3}+\alpha^{1/6}+B(\alpha, p_0))$.  This relation can lead to the Lindblad-type equation, 
\begin{align}
\label{eq:modifLind2}
\dot{\rho} =-i[H,\rho(t)]-A[H^{3/2},[H^{3/2},\rho(t)]],
\end{align}
in which case,
\begin{align}
    A=(t_pl_p^2)^{2/3}.
\end{align}
Here we show different power-law dependence of the decoherence parameters, which are usually studied in a model-independent manner~\cite{Stuttard:2020qfv}, can in principle arise from different variants of GUP.
Here, it is to be noted that the exact equations and the exact expression of the decoherence parameters depends on the entire form of the GUP. Here we have just tried to obtain the power law behaviour and the order of the decoherence parameters.

\section{Results}
\label{sec:results}
Decoherence effects in neutrino oscillations due to quantum gravity have been studied from model-independent perspective in~\cite{Klapdor-Kleingrothaus:2000kdx, Stuttard:2020qfv,Stuttard:2021uyw}. In this work, we have used a specific model where the effect of quantum gravity manifests in the LQGUP. In this section, we try to compare the consequences of decoherence due to the LQGUP, as well as possible extensions, to that of the existing model-independent and other experimental searches of neutrino decoherence.

\subsection{LQGUP Case}
\label{sbsec:lqgupcase}

\noindent
\textbf{Decoherence parameter}:
In general model-independent studies, the energy-dependent decoherence parameter, $\Gamma$ is expressed as a power law,
\begin{equation}
\Gamma(E) = \Gamma(E_0)\left(\frac{E}{E_0}\right)^n,
\end{equation}
where $E_0$ is the reference energy, which we take to be $\mathcal{O}$(GeV), and $n$ is the power.
As we can see from subsection~\ref{sbsec:3flav}, in this study we have three different decoherence parameters. These decoherence parameters are given as follows,
\begin{subequations}
\begin{align}
\Gamma_{21}(E) &= \frac{3}{16}(t_Pl_P^2)(\Delta m_{21}^2)^4 \left(\frac{E}{\text{GeV}}\right)^{-4},\\
\Gamma_{31}(E) &= \frac{3}{16}(t_Pl_P^2)(\Delta m_{31}^2)^4 \left(\frac{E}{\text{GeV}}\right)^{-4},\\
\Gamma_{32}(E) &= \frac{3}{16}(t_Pl_P^2)(\Delta m_{32}^2)^2(\Delta m_{31}^2+\Delta m_{21}^2)^2 \left(\frac{E}{\text{GeV}}\right)^{-4},
\end{align}
\label{eq:decohParam}
\end{subequations}
where $t_p$ and $l_p$ are Planck time and length, respectively.
From this, we can note that in this model, the power $n=-4  $.
\begin{figure}[H]
\centering
\subfloat[]{{\includegraphics[scale=0.5]{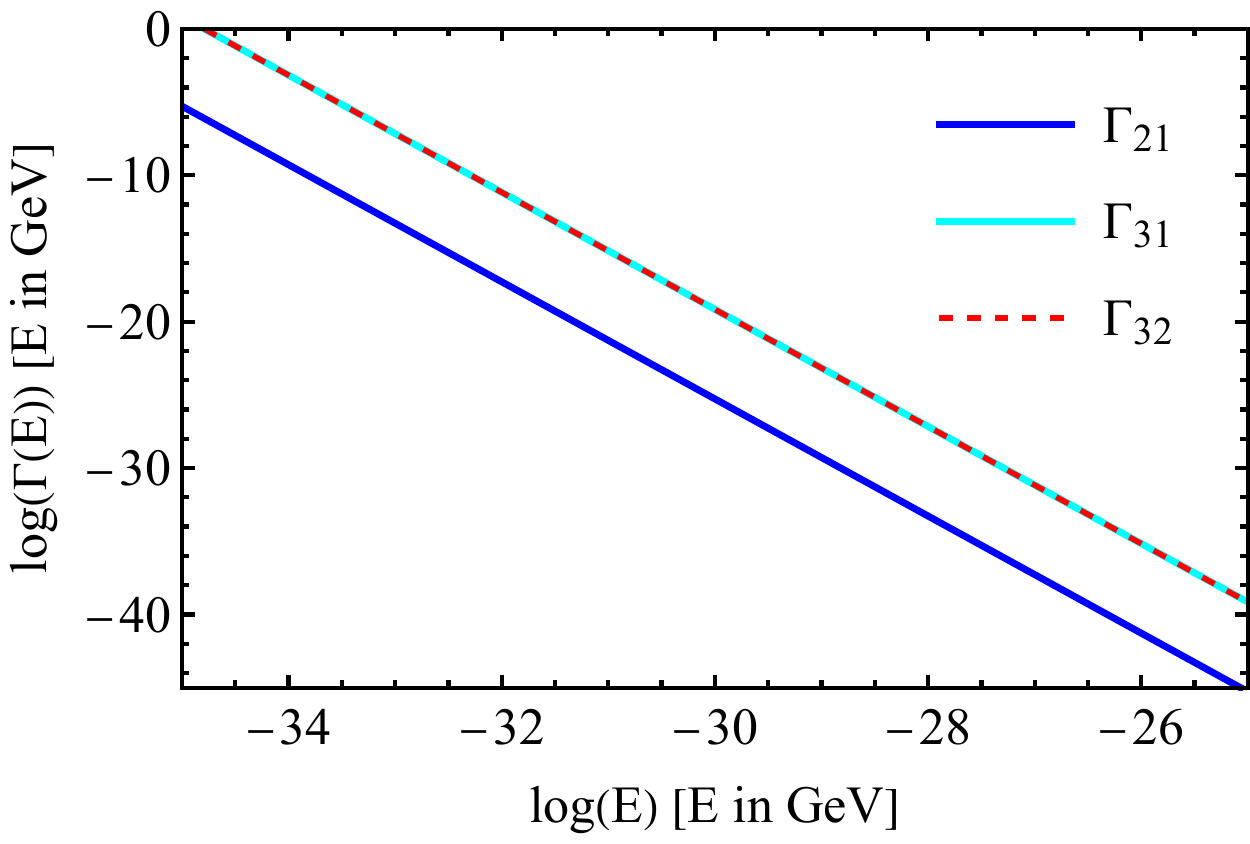}}  \label{fig:fig1a}}
\quad \quad
\subfloat[]{{\includegraphics[scale=0.5]{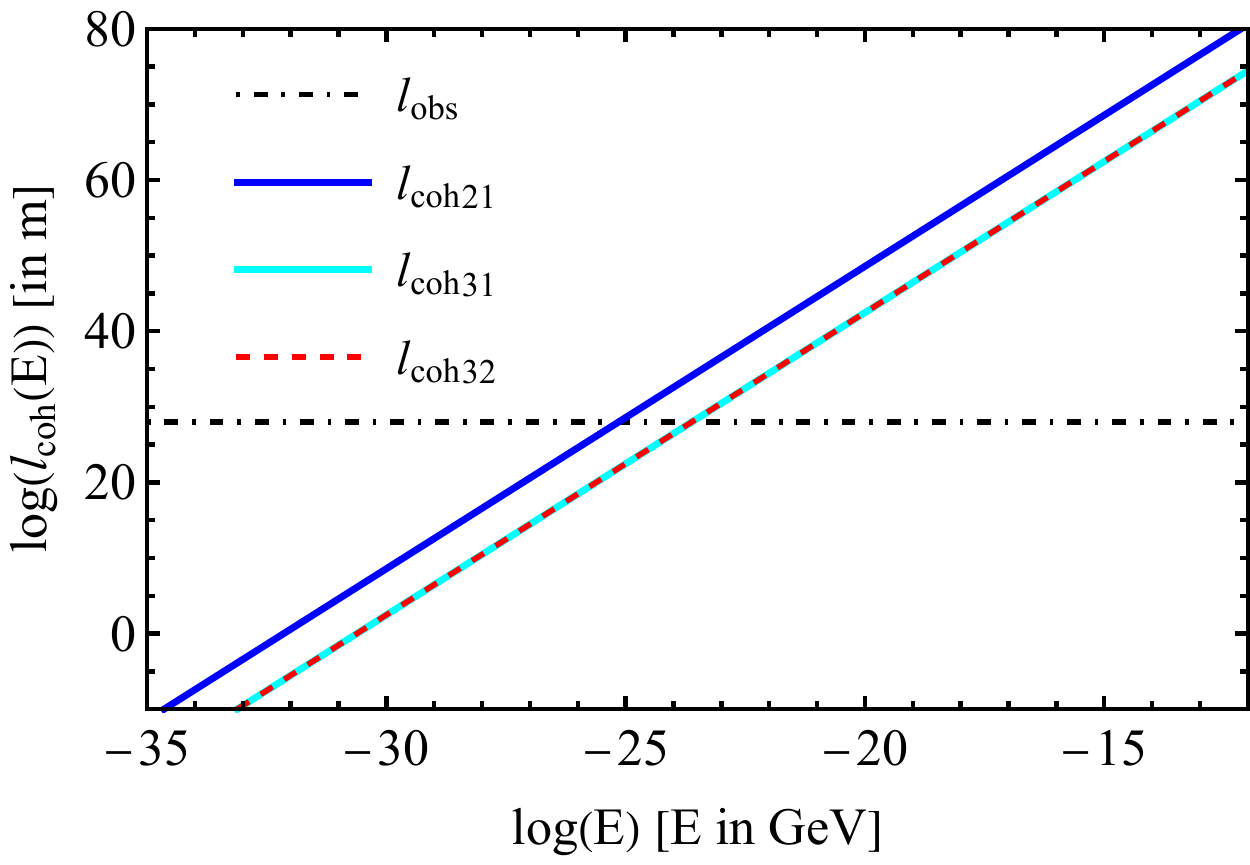}} \label{fig:fig1b}}
\caption{The dependence on energy for (a) the decoherence parameter and (b) the coherence length in the LQGUP scenario.}
\label{Fig1}
\end{figure} 
Apart from this we can see that unlike some phenomenological models~\cite{Stuttard:2020qfv}, there are multiple decoherence parameters with different expressions depending on the mass-squared differences of the neutrino masses. In Figure (\ref{fig:fig1a}), the dependence of these decoherence parameters with respect to the energy are shown. It is to be noted that $\Gamma_{31}(E)$ and $\Gamma_{32}(E)$ are very close in the given range while $\Gamma_{21}(E)$ differs by a small amount. The underlying reason for this is the numerical values of the mass-squared differences which are $\Delta m_{21}^2 \sim 10^{-5}$ eV$^2$, $\Delta m_{31}^2 \sim 10^{-3}$ eV$^2$ and $\Delta m_{32}^2 \sim 10^{-3}$ eV$^2$. 

\noindent
\textbf{Coherence length:}
The coherence length $l_{\rm coh}$ is defined as the propagation length after which the amplitude of the survival and the transition probabilities of the neutrinos are diminished by a factor of $e$, i.e.,
\begin{equation}
l_{\rm coh}=\frac{1}{\Gamma (E)}.
\end{equation}
Thus in our case, the coherence length increases rapidly with energy since $l_{\rm coh} \propto E^4$, as is evident from Eq.~\eqref{eq:decohParam}. This suggests that in this model, the effects of the quantum gravity becomes more prominent on lower energy regimes as compared to higher energy regimes of the neutrinos. 
The coherence length can also be thought of as the interaction mean free path which is the average distance a neutrino covers before interacting with a virtual black hole (VBH) in the space-time foam~\cite{Stuttard:2020qfv}. Therefore, it is likely that in this model, the high energy neutrinos are much less likely to interact with a VBH as compared to the low energy neutrinos as opposed to the models where $n>0$~\cite{Stuttard:2020qfv}. It is also possible that an interaction with the VBH results in more noticeable consequences for low energy neutrinos in comparison with high energy neutrinos. An effect of this kind can also arise from some other model of quantum gravity where space-time uncertainties at very small length scales results in lightcone fluctuations.~\cite{Wheeler:1955zz}. It has been shown in \cite{Stuttard:2021uyw} that lightcone fluctuations can also cause decoherence effects in neutrino oscillations in vacuum. An effect of that kind can lead to an energy dependence on the decoherence parameter $\propto E^{2(n-1)}$ where $n$ denotes the dependence of the distance uncertainty on energy. Therefore, a model with $n=-1$ can lead to an energy dependence of the decoherence effects similar to the one obtained in this study through the LQGUP model. It is also of interest that the decoherence parameters dependent on different frequencies, i.e. different mass splittings ($\Delta m_{21}^2$, $\Delta m_{31}^2$ and $\Delta m_{32}^2$) are a common feature in~\cite{Stuttard:2021uyw} as well.  Although the exact form of the decoherence parameter depends on the other variables in the expression of distance fluctuation used in \cite{Stuttard:2021uyw}, it is a possibility that the lightcone fluctuations can be the underlying reason of the decoherence effects arising from the LQGUP. In Figure~(\ref{fig:fig1b}) the dependence of decoherence parameters on the neutrino energies are shown. It is to be noted that $l_{\text{coh}31}$ and $l_{\text{coh}32}$ has similar values for the given energy range whereas $l_{\text{coh}21}$ differs slightly which is again due to the hierarchy in the mass-squared differences. The length corresponding to the size of the observable Universe ($\sim 10^{28}$m~\cite{Kolb:1990vq}) is shown by the horizontal dash-dotted line. This corresponds to the fact that according to this model, neutrinos with energy $\mathcal{O}(10^{-23})$ GeV or lower has the coherence length less than that of the observable universe. This implies that the decoherence effect due to the GUP model is predominant for the lower energy neutrinos.
%

%

\subsection{Possible extensions}
\label{sbsec:possibleresult}
In section~\ref{sec:extn}, we have shown how for a different power-law the Lindblad-type equation can be modified through the modification in the Heisenberg algebra. The Eqs.~\eqref{eq:modifLind}  and \eqref{eq:modifLind2} can now be used instead of Eq.~\eqref{eq:modlindbladgen} to calculate the survival and transition probabilities of the neutrinos. It is to be noted that for the case where $n=-2$, the Lindblad operators $D_a \propto H$ and for the case where $n=-3$, the Lindblad operators $D_a \propto H^{3/2}$. 
\noindent
\textbf{Decoherence parameter}: The decoherence which will take place for the propagation of neutrinos through vacuum in this case will be governed by Eq.~\eqref{eq:modifLind} and this leads to a decoherence parameters with $n=-2$ power-law,
\begin{subequations}
\begin{align}
\Gamma_{21}(E) = \frac{3}{4}( t_p l_p^2)^{1/3}(\Delta m^2_{21})^2 \left(\frac{E}{\text{GeV}}\right)^{-2},\\
\Gamma_{31}(E) = \frac{3}{4}( t_p l_p^2)^{1/3}(\Delta m^2_{31})^2 \left(\frac{E}{\text{GeV}}\right)^{-2},\\
\Gamma_{32}(E) = \frac{3}{4}( t_p l_p^2)^{1/3}(\Delta m^2_{32})^2 \left(\frac{E}{\text{GeV}}\right)^{-2}.
\end{align}
\label{eq:decohParam2}
\end{subequations}
\begin{figure}[t]
\centering
\subfloat[]{{\includegraphics[scale=0.77]{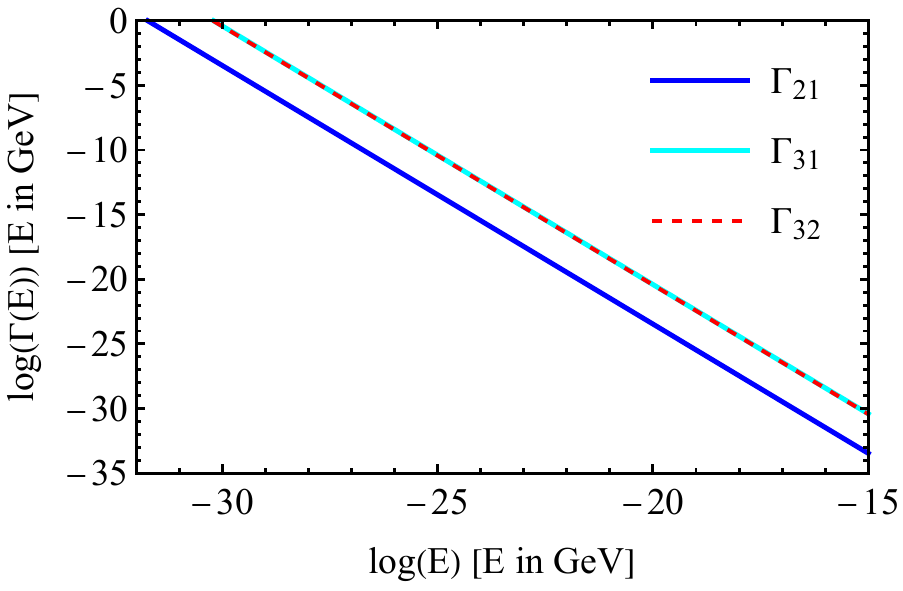}}  \label{fig:fig2a}}
\quad \quad
\subfloat[]{{\includegraphics[scale=0.75]{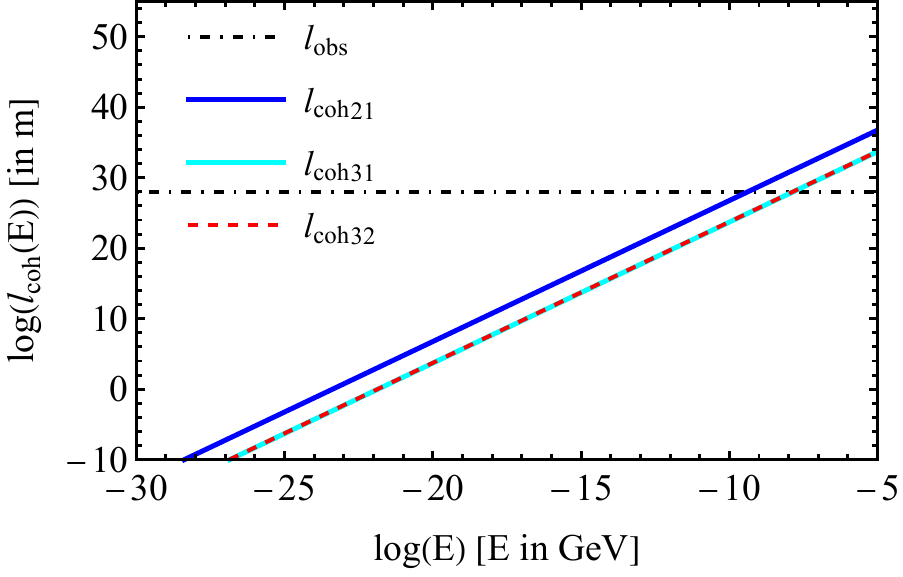}} \label{fig:fig2b}}
\caption{The dependence on energy for (a) the decoherence parameter and (b) the coherence length in the extended GUP scenario with $n=-2$.}
\label{Fig2}
\end{figure} 
Similarly, for the $n=-3$ power law, the decoherence parameters are approximately of the form,
\begin{subequations}
\begin{align}
\Gamma_{21}(E) &=  \frac{3}{8}( t_p l_p^2)^{2/3}(\Delta m^2_{21})^3 \left(\frac{E}{\text{GeV}}\right)^{-3},\\
\Gamma_{31}(E) &=  \frac{3}{8}( t_p l_p^2)^{2/3}(\Delta m^2_{31})^3 \left(\frac{E}{\text{GeV}}\right)^{-3},\\
\Gamma_{32}(E) &=  \frac{3}{8}( t_p l_p^2)^{2/3}\left((\Delta m_{31}^{2})^{3/2}-(\Delta m_{21}^{2})^{3/2}\right)^2\left(\frac{E}{\text{GeV}}\right)^{-3}.
\end{align}
\label{eq:decohParam3}
\end{subequations}
\begin{figure}[t]
\centering
\subfloat[]{{\includegraphics[scale=0.77]{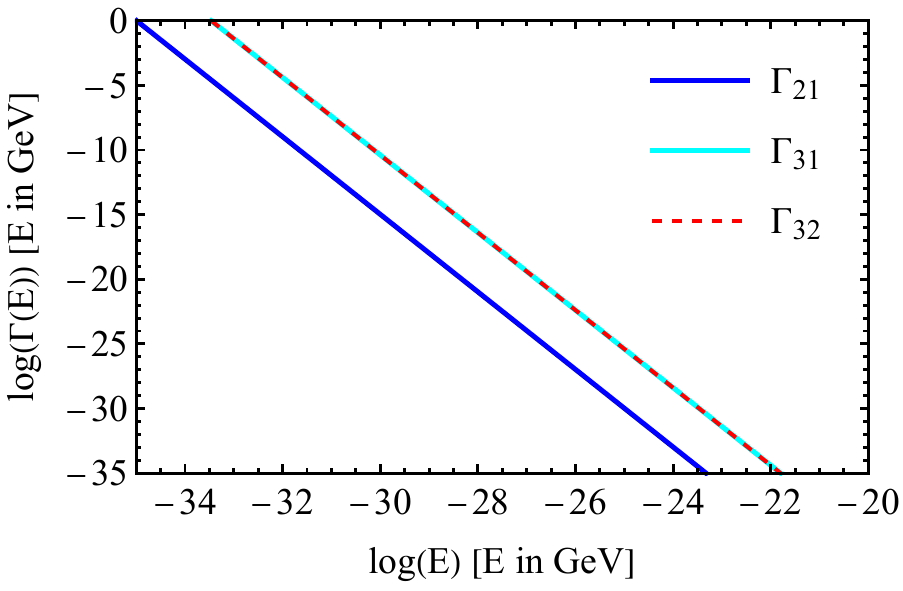}}  \label{fig:fig3a}}
\quad \quad
\subfloat[]{{\includegraphics[scale=0.75]{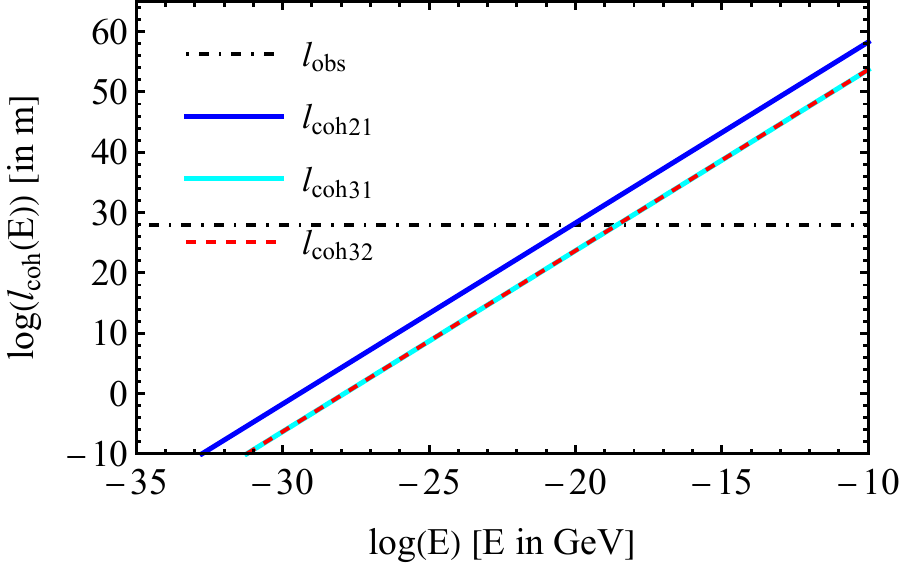}} \label{fig:fig3b}}
\caption{The dependence on energy for (a) the decoherence parameter and (b) the coherence length in the extended GUP scenario with $n=-3$.}
\label{Fig3}
\end{figure} 
In figures~(\ref{fig:fig2a}) and (\ref{fig:fig3a}), we show the dependence of these decoherence parameters with respect to the energy. In comparison to the $n=-4$ case, the values of the decoherence parameters are much larger for the same energy range of the neutrinos. But similar to the previous scenario, this extension also suggests higher decoherence for lower energy neutrinos.
\textbf{Coherence length}: The coherence length, as mentioned previously, $l_{\rm_{coh}}=1/\Gamma(E)$. For the cases with $n=-2$ and $n=-3$ the coherence lengths are proportional to $E^2$ and $E^3$ respectively. In figure~(\ref{fig:fig2b}) and (\ref{fig:fig3b}), we show the dependence of the coherence lengths on the energy of the neutrinos. In this case also, higher energy neutrinos will have longer coherence length than that of lower energy neutrinos, but this model predicts a shorter coherence length than the previous model for the same energy neutrinos.
In the scenario where $n=-2$, for lower energy neutrinos e.g., for solar neutrinos with energy $\mathcal{O}(1)$ MeV~\cite{Giunti:2007ry} the decoherence parameter $\Gamma \sim 10^{-55}$ GeV and the coherence length $l_{\rm_{coh}} \sim 10^{39}$ m. However, for the cosmic neutrino background (CNB) neutrinos with energy $\mathcal{O}(10^{-15})$ GeV~\cite{Giunti:2007ry} the decoherence parameter $\Gamma \sim 10^{-31}$ GeV and the coherence length $l_{\rm_{coh}} \sim 10^{15}$ m. Clearly, this is in the paradigm of future experiments where finer details of the CNB can be measured.

\subsection{Effects on free-streaming neutrinos}
\label{sbsec:freestreamingresult}
Neutrinos decoupled from the primordial plasma at $T\sim 1.3$ MeV and they were of the free-streaming nature due to their ultra-relativistic nature~\cite{Giunti:2007ry}. After decoupling, the temperature of these neutrinos is inversely proportional to the scale factor of the universe($T_{\nu}\propto 1/a(t)$). The decoupling occurred in the radiation dominated era. Since then, the universe went through matter- and cosmological constant-dominated era respectively. In all these different eras, the scale factor, $a(t)$ had different dependencies on time, i.e. during the radiation dominated era, $a(t)\propto t^{1/2}$, during the matter dominated era, $a(t)\propto t^{2/3}$ and during the cosmological constant dominated era $a(t)\propto e^{Ht}$ where $H$ is the Hubble constant~\cite{Giunti:2007ry}. Therefore, in all these different eras, the temperature and therefore the energy of these neutrinos will have different dependencies on the propagation length as the propagation length for neutrinos, $L\sim t$. Therefore, for all practical purposes, during the radiation dominated era, the energy of the relic neutrinos, $E\propto L^{-1/2}$, during the matter dominated era, $E\propto L^{-2/3}$, and during the cosmological constant dominated era, $E\propto e^{-H_0L}$ where $H_0$ is the Hubble constant today.
According to the standard picture of neutrino oscillation, the flavour ratio of these relic neutrinos is $1:1:1$~\cite{PTOLEMY:2019hkd}. We expect that in the presence of decoherence due to interactions with the vacuum, the relic neutrino flavour ratio might differ. To check this, we have used the dependencies of $E$ on the propagation length $L$ during different eras, mentioned in the previous paragraph, in the expressions of transition and survival probabilities, such as Eqs.~\eqref{eq:Pee3by3} and~\eqref{eq:Pemu3by3}. We find  that for the LQGUP ($n=-4$) case and the possible extension where $n=-3$, the flavour ratio today ($t\sim 14$~Ga) remains $1:1:1$ for the different flavour of neutrinos. For the case where $n=-2$, the flavour ratio changes by $\mathcal{O}(10^{-16})$.
In upcoming experiments, e.g., PTOLEMY, the expected number of detection of $\nu_e$ is $\mathcal{O}(10)$ per year~\cite{PTOLEMY:2019hkd}. Hence, it is not possible to detect a modification of $\mathcal{O}(10^{-16})$ through an experiment with the working precision of that of PTOLEMY. However, the situation will not be so bleak in case of GUP models predicting $n>0$, a detailed study of which we leave for a future work.

\section{Summary and Conclusion}
\label{sec:concl}
The LQGUP model is shown to give rise to a decoherence in the neutrino oscillation with decoherence parameter having a $n=-4$ power-law dependence on energy. Therefore, lower energy neutrinos will experience significantly more effect of the decoherence than the high energy neutrinos for a given propagation length. This strong quartic suppression of the decoherence effect with energy occurs because the Lindblad operators arising due to the LQGUP are proportional to the square of the Hamiltonian which in this case is inversely proportional to $E$. Through this approach, the mean free path of the neutrinos can be estimated which gives us the idea regarding the structure of the vacuum in the quantum gravity theories which predicts the LQGUP. Moreover, the decoherence of the neutrinos depends on the energy and therefore the vacuum interacts differently for different energy ranges of the neutrinos, e.g., for CNB neutrinos of energy $\mathcal{O}(10^{-15})$ GeV the mean free path is $\mathcal{O}(10^{63})$ m which suggests that a neutrino in this energy range will on average interact with the vacuum in way that diminishes its transition and survival probabilities by a factor of $e$ after every $\mathcal{O}(10^{63})$ m of propagation. One of the possible ways this vacuum interaction can occur is via VBH. Another possibility can be the lightcone fluctuations. As the energy goes down, the mean free path also reduces rapidly which suggests that lower energy neutrinos will experience more decoherence than higher energy neutrinos in LQGUP model. Since, the coherence length for even CNB neutrinos are very much larger than the size of the observable universe, the LQGUP predictions are impossible to be detected through existing experimental facilities for any natural source of neutrino.
In principle, one can use similar treatments to get other power-law dependencies from different variants of GUP. As examples we have shown the cases of $n=-2$ and $n=-3$. There can be different forms of GUPs which can give rise to modified momentum relations and consequently lead to a decoherence parameters with $n=-2$ and $n=-3$ power-law dependence on energy. 
In this case also, the decoherence effects are dominant in case of low energy neutrinos but unlike the LQGUP model, for the case of $n=-2$, the CNB neutrinos will have a decoherence length $\mathcal{O}(10^{15})$ m which is very much realistic for the purposes of future detection. 
As far as cosmological observations are concerned, decoherence can, in principle, affect the free streaming properties of the relic neutrinos in the CNB. We find that though decoherence effects lead to no change in the flavour ratio of the CNB for the case of $n=-4$ and $n=-3$, but for $n=-2$ it will lead to a change of $\mathcal{O}(10^{-16})$ in the flavour ratio of the CNB. But, due to the well known difficulties in direct detection of the CNB, this difference is too insignificant to be detected through the upcoming experiments in the near future.

This result also serves as a motivation to study the different variants of GUP which originates from different theories of quantum gravity and perform a rigorous analysis to set the stage for a probe of quantum gravity from the neutrino oscillation experiment.
This study connects a quantum gravitational modification in the uncertainty principle to the energy dependence of the interaction of neutrinos with the vacuum. If a GUP model predicts that the Lindblad operators $D_a$ are proportional to $H^{-n}$ then that corresponds to the fact that the decoherence parameter $\Gamma (E)$ is proportional to $E^{2n}$. This kind of correlation can be used to pick a particular form of GUP which supports the neutrino oscillation and decoherence due to the interaction with the vacuum from the available data. Such model-dependent work can be used in future to indirectly test the claims of different models of quantum gravity. Similar methods can be followed to investigated other GUP models as well. 
For GUP models which gives rise to neutrino decoherence with $n>0$, the flavour ratio of high energy neutrinos from different natural and artificial sources could be investigated in order to test those GUP models. Such model-dependent work could confirm, rule out or put stringent bounds on different GUP models and indirectly test the claims of the different quantum gravity models which lead to these GUPs. Similarly, decoherence effects on different sources of high energy neutrinos will be optimized for different models of GUP with positive values of $n$ and one could indirectly test the quantum gravity models through the similar decoherence studies.
A generalised GUP model can also be devised using the general power law energy dependence of the decoherence parameter which will provide key insights into the relationship between energy dependence of the decoherence parameter and the modifications to the Heisenberg uncertainty principle.


%
 
\section*{Acknowledgments}
We thank Victor Mukherjee for useful discussions. UKD acknowledges the support from Department of Science and Technology (DST), Government of India under the Grant Reference No. SRG/2020/000283.  UKD also acknowledges  hospitality of APCTP Pohang towards the completion of this project.

\bibliographystyle{JHEP}
\bibliography{gup_nuDecoh_ref.bib}

\end{document}